\documentclass[12pt]{report}
\usepackage{epsfig}
\usepackage{amsbsy,amssymb}
\makeatletter\renewcommand\@eqnnum{[\theequation]}\makeatother
\begin{document}
\font\mybigsf=cmssbx10 at 17pt
\font\mysfb=cmssbx10 at 14pt
\font\mysfi=cmssi10 at 14pt
\font\mysfbs=cmssbx10
\font\mysf=cmss10
\baselineskip=20pt
\noindent{\mybigsf YANG--BAXTER EQUATIONS}

\noindent{\mysfbs Jacques H.H.\ Perk \&\ Helen Au-Yang},
{\mysf Department of Physics,
Oklahoma State University, Stillwater, OK 74078-3072, USA}

\bigskip\baselineskip=14.5pt
\noindent{\mysfb Introduction}
\par\noindent
The term Yang--Baxter Equations (YBE) has been coined by Faddeev in the late
1970s to denote a principle of integrability, i.e.\ exact solvability, in a
wide variety of fields in physics and mathematics. Since then it has become
a common name for several classes of local equivalence transformations in
statistical mechanics, quantum field theory, differential equations, knot
theory, quantum groups, and other disciplines. We shall cover the various
versions and their relationships, paying attention also to the early
historical development.

\goodbreak\bigskip
\noindent{\mysfi Electric networks}
\par\noindent
The first such transformation came up as early as 1899 when the Brooklyn
engineer Kennelly published a short paper, entitled {\em the equivalence of
triangles and three-pointed stars in conducting networks}.
\begin{figure}[tbh]
\begin{center}
\epsfig{file=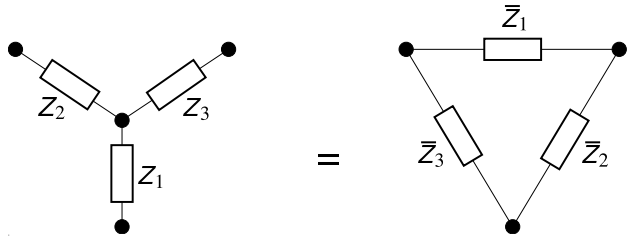,width=0.6\hsize}
\end{center}
\caption{Star-triangle equation for impedances.}
\label{fig1}
\end{figure}
This work gave the definite answer to such questions as whether it is better
to have the three coils in a dynamo---or three resistors in a
 network---arranged as a star or as a triangle, see figure \ref{fig1}. Using
Kirchhoff's laws, the two situations in figure \ref{fig1} can be shown to be
equivalent provided
\begin{eqnarray}
&&Z_1\overline Z_1=Z_2\overline Z_2=Z_3\overline Z_3\nonumber\\
&&\qquad=Z_1Z_2+Z_2Z_3+Z_3Z_1
\label{eq1}\\
&&\qquad=\overline Z_1\overline Z_2\overline Z_3/
(\overline Z_1+\overline Z_2+\overline Z_3)
\label{eq2}
\end{eqnarray}
Here one has to take either [\ref{eq1}] or [\ref{eq2}] as second line of
the equation, depending which direction the transformation is to go. The
{\em star-triangle transformation} thus defined is also known under other
names within the electric network theory literature as wye-delta
($\hbox{Y}-\Delta$), upsilon-delta ($\Upsilon-\Delta$), or tau-pi
($\hbox{T}-\Pi$) transformation.

\goodbreak\bigskip
\noindent{\mysfi Spin models}
\par\noindent
When Onsager wrote his monumental paper on the Ising model published in 1944,
he made a brief remark on {\em an obvious star-triangle transformation}
relating the model on the honeycomb lattice with the one on the triangular
lattice. His details on this were first presented in Wannier's review article
of 1945. However, the star-triangle transformation played a much more crucial
role in Onsager's reasoning, as it is also intimately connected with his
elliptic function uniformizing parametrization.

Furthermore, it implies the commutation of transfer matrices and spin-chain
hamiltonians. Only in his Battelle lecture of 1970 did Onsager explain how he
used this remarkable observation in his derivation of the formula for the
spontaneous magnetization which he had announced as a conference remark in
1948 and of which the first complete derivation had been published by Yang in
1952 using a completely different method.

\begin{figure}[tbh]
\begin{center}
\epsfig{file=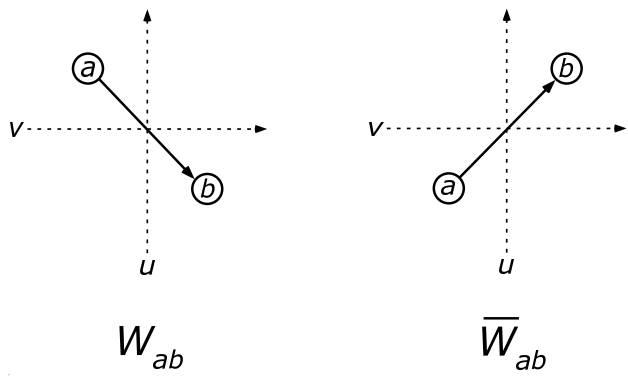,width=0.6\hsize}
\end{center}
\caption{Spin model weights $W_{ab}(p,q)$ and $\overline W_{ab}(p,q)$.}
\label{fig2}
\end{figure}

Many other applications and generalizations have since appeared. Most
generally, we can consider a system whose state variables---also called
spins---take values from some suitable discrete or continuous sets. The
interactions between spins $a$ and $b$ are given in terms of weight factors
$W_{ab}$ and $\overline W_{ab}$, which are complex numbers in general, see
figure \ref{fig2}. One quantity of special interest is the partition
function---sum of the product of all weight factors over all allowed spin
values. The integrability of the model is expressed by the existence of
spectral variables---rapidities $p,q,r,\ldots$---that live on oriented lines,
two of which
cross between $a$ and $b$ as indicated by the dotted lines in figure
\ref{fig2}. Arrows from $a$ to $b$ are added to keep track of the ordering
of $a$ and $b$ in case the weights are chiral (not symmetric).

In Onsager's special Ising model case the spins take values
$a,b,c,\ldots=\pm1$ and the weight factors are the usual real positive
Boltzmann weights depending on the product $ab=\pm1$, uniformizing
variable $p-q$ and elliptic modulus $k$. In the integrable chiral Potts model
the weights depend on $a-b$ mod $N$, with $a,b=1,\ldots,N$, whereas the
rapidities $p$ and $q$ are living in general on a higher-genus curve.

When the weights are asymmetric in the spins, there are two sets of
star-triangle equations which can be expressed both pictorially---see figure
\ref{fig3}---and algebraically:
\begin{figure}[tbh]
\begin{center}
\epsfig{file=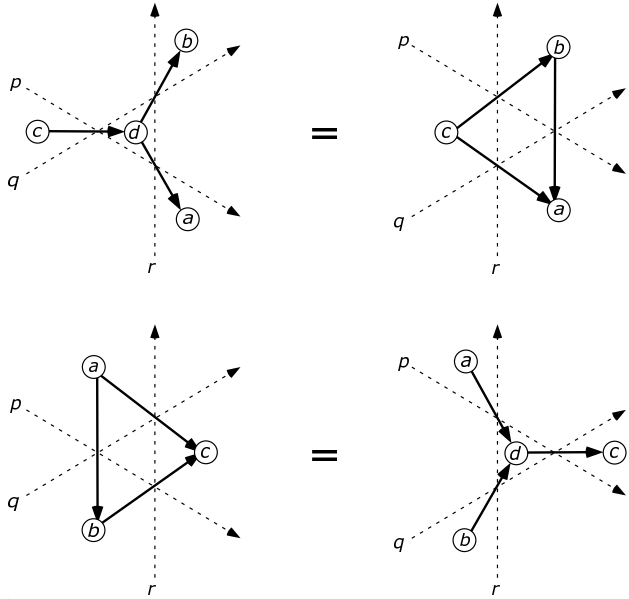,width=0.6\hsize}
\end{center}
\caption{Star-triangle equation.}
\label{fig3}
\end{figure}
\begin{eqnarray}
&&\sum_{d}\,
\overline W{}_{cd}(p,q)\overline W{}_{db}(q,r)W{}_{da}(p,r)\nonumber\\
&&\qquad=R(p,q,r)\,
W{}_{ba}(p,q)W{}_{ca}(q,r)\overline W{}_{cb}(p,r)
\label{eq3}\\ \cr
&&\overline R(p,q,r)\,
W{}_{ab}(p,q)W{}_{ac}(q,r)\overline W{}_{bc}(p,r)\nonumber\\
&&\qquad=\sum_{d}\,
\overline W{}_{dc}(p,q)\overline W{}_{bd}(q,r)W{}_{ad}(p,r)
\label{eq4}
\end{eqnarray}
Note that eqs.\ [\ref{eq3}] and [\ref{eq4}] differ from each other by the
transposition of both spin variables in all six weight factors. In general
there may also appear scalar factors $R(p,q,r)$ and $\overline R(p,q,r)$,
which can often be eliminated by a suitable renormalization of the weights.
If $a$, $b$, and $c$ take values in the same set, we can sum over
$a=b=c$ showing that $R=\overline R$ in that case.

The Kennelly star-triangle equation [\ref{eq1}], [\ref{eq2}] can be recovered
as a special limit of a spin model where the states are continuous variables.

\goodbreak\bigskip
\noindent{\mysfi Knot theory and braid group}
\par
\begin{figure}[tbh]
\begin{center}
\epsfig{file=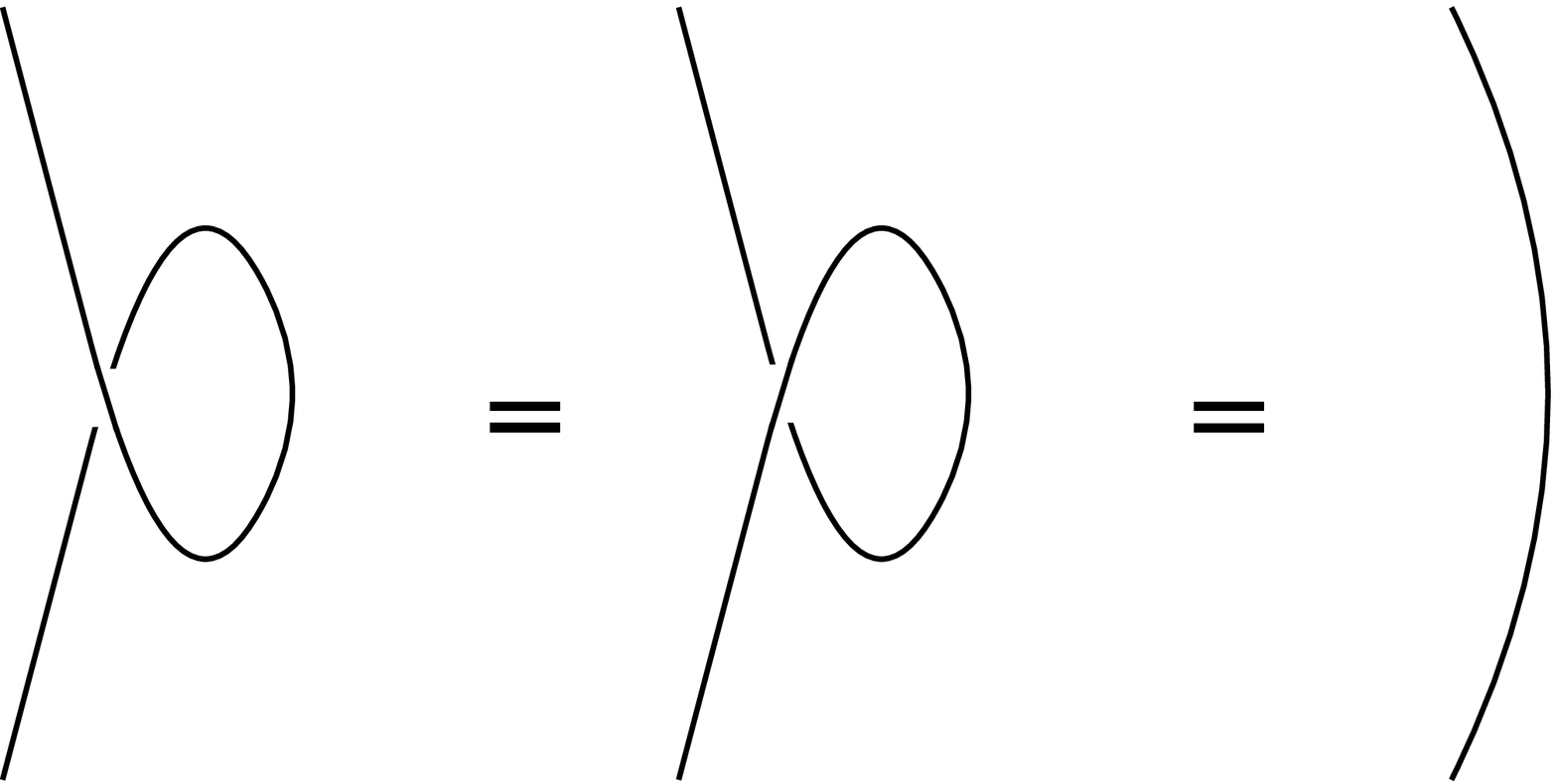,width=0.25\hsize}\hspace{0.12\hsize}
\epsfig{file=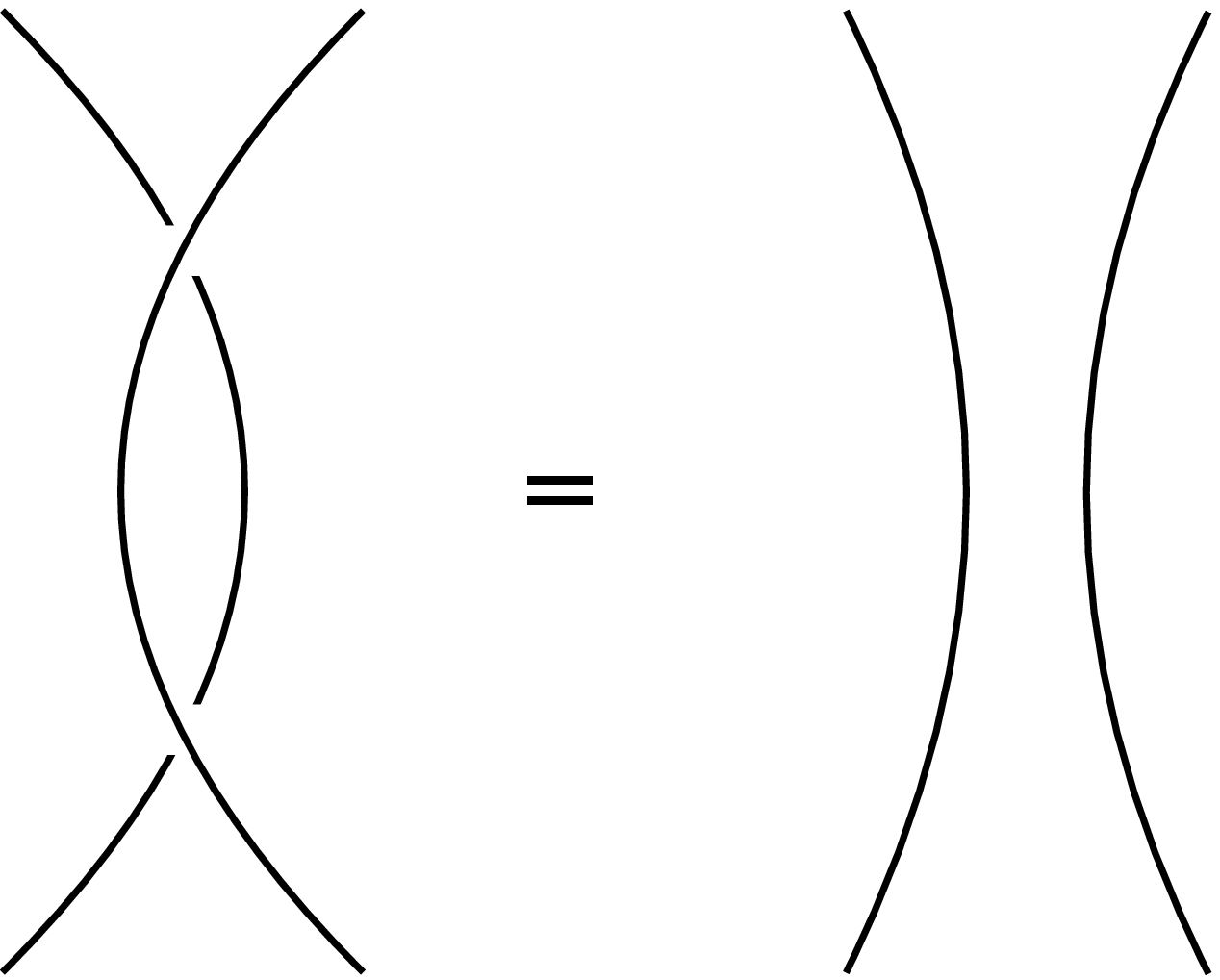,width=0.18\hsize}
\end{center}
\vspace{1em}
\begin{center}
\epsfig{file=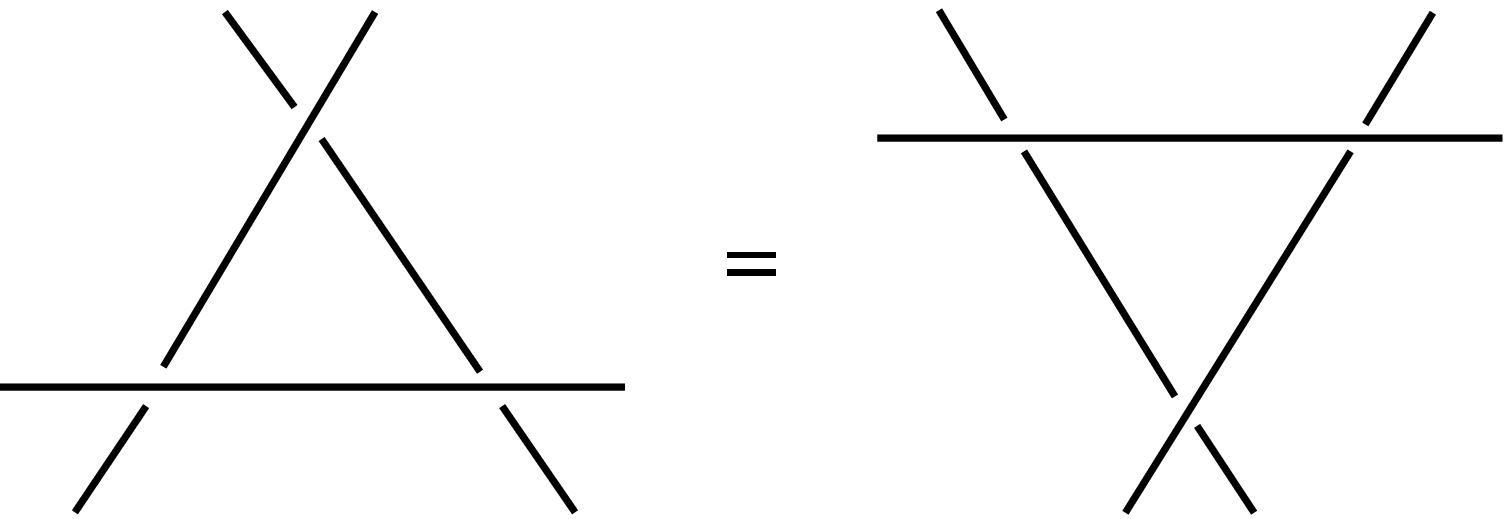,width=0.5\hsize}
\end{center}
\caption{Reidemeister moves of type I, II, and III.}
\label{fig4}
\end{figure}

\noindent
A seemingly totally different situation occurs in the theory of knots, links,
tangles, and braids. In 1926 Reidemeister showed that only three types of
moves suffice to show the equivalence between two different
configurations, see figure \ref{fig4}. Moves of type I---removing simple
loops---do
not apply to braids. Moves of type II, for which one strand crosses twice
over another strand, can be reformulated for braids, namely that an
over-crossing is the inverse of an under-crossing. The Reidemeister move of
type III is a precursor of the more general Yang--Baxter moves and can be
represented also by the defining relations of Artin's braid group. Let
$\mathsf{R}_{i,i+1}$ be the operator representing the situation in which the
strand in position $i$ crosses over the one in position $i+1$. Then a braid
can be represented by a product of $\mathsf{R}_{j,j+1}$'s and their inverses,
provided
\begin{eqnarray}
\mathsf{R}_{i,i+1}\mathsf{R}_{i+1,i+2}\mathsf{R}_{i,i+1}
=\mathsf{R}_{i+1,i+2}\mathsf{R}_{i,i+1}\mathsf{R}_{i+1,i+2}
\label{eq5}
\end{eqnarray}
and
\begin{equation}
[\mathsf{R}_{i,i+1},\mathsf{R}_{j,j+1}]=0,
\quad\hbox{if}\quad\vert i-j\vert\ge 2
\label{eq6}
\end{equation}
and similar relations in which $\mathsf{R}_{i,i+1}$ and/or
$\mathsf{R}_{i+1,i+2}$ are replaced by their inverses.

\goodbreak\bigskip
\noindent{\mysfi Factorizable S-matrices and Bethe Ansatz}
\par
\begin{figure}[tbh]
\begin{center}
\epsfig{file=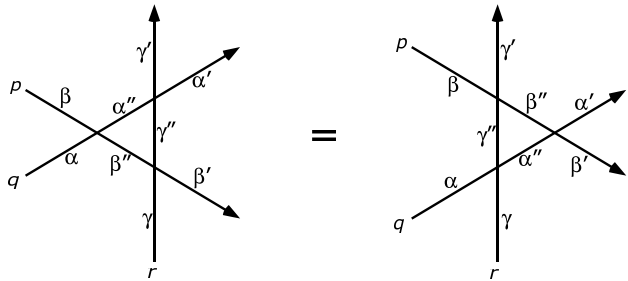,width=0.6\hsize}
\end{center}
\caption{Vertex model Yang--Baxter equation.}
\label{fig5}
\end{figure}

\noindent
In the early 1960s Lieb and Liniger solved the one-dimensional Bose gas with
delta-function interaction using the Bethe Ansatz. Yang and McGuire then
tried to generalize this result to systems with internal degrees of freedom
and to fermions. This led to the discovery of the condition for factorizable
S-matrices by McGuire in 1964, represented pictorially by figure \ref{fig5}.
Here the worldlines of the particles
are given. Upon collisions the particles can only exchange their rapidities
$p,q,r$, so that there is no dispersion. Also indicated are the internal
degrees of freedom in Greek letters. In other words, the three-body 
$S$-matrix can be factorized in terms of two-body contributions and the
order of the collisions does not affect the final outcome. McGuire also
realized that this  condition is all you need for the consistency of
factoring the $n$-body $S$-matrix in terms of two-body $S$ matrices.
The consistency condition is obviously related to the Reidemeister move
of type III in figure \ref{fig4}.

Yang succeeded in solving the spin-$\frac12$ fermionic model using a
nested Bethe Ansatz, utilizing a generalization of Artin's braid relations
[\ref{eq5}] and [\ref{eq6}],
\begin{eqnarray}
&&\check{\mathsf{R}}_{i,i+1}(p-q)\check{\mathsf{R}}_{i+1,i+2}(p-r)
\check{\mathsf{R}}_{i,i+1}(q-r)\cr
&&\quad=\check{\mathsf{R}}_{i+1,i+2}(q-r)\check{\mathsf{R}}_{i,i+1}(p-r)
\check{\mathsf{R}}_{i+1,i+2}(p-q)
\label{eq7}
\end{eqnarray}
He submitted his findings in two short papers in 1967.
The $\check{\mathsf{R}}$ operators in eq.\ [\ref{eq7}]---a notation introduced
later by the Leningrad school---depend on differences of two momenta or two
relativistic rapidities. Sutherland solved the general spin case using
repeated nested Bethe Ans\"atze, while Lieb and Wu used Yang's work to solve
the one-dimensional Hubbard model.

\goodbreak\bigskip
\noindent{\mysfi Vertex models}
\par\noindent
Since Lieb's solution of the ice model by a Bethe Ansatz there have been many
developments on vertex models, in which the state variables live on line
\begin{figure}[tbh]
\begin{center}
\epsfig{file=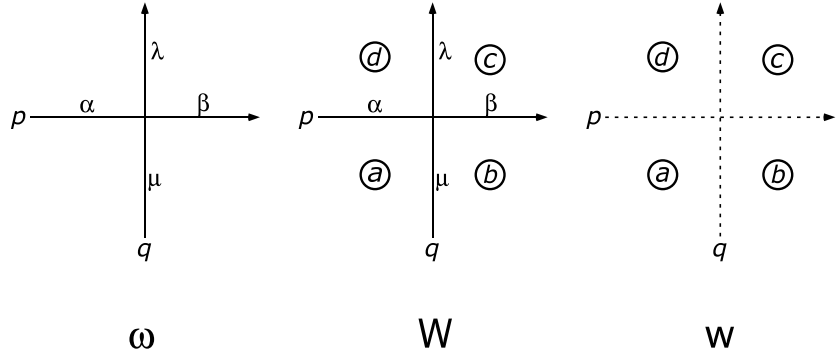,width=0.8\hsize}
\end{center}
\caption{Vertex model weight
$\omega{}^{\lambda\beta}_{\alpha\mu}(p,q)$,
mixed model weight
$W{}^{\lambda\beta}_{\alpha\mu}|{}^{dc}_{ab}(p,q)$
and IRF model weight
$w{}^{dc}_{ab}(p,q)$.}
\label{fig6}
\end{figure}
segments and a weight factors $\omega{}^{\lambda\beta}_{\alpha\mu}$ is
assigned to a vertex where four line segments with the four states
$\alpha,\mu,\lambda,\beta$ on them meet, see figure \ref{fig6}.

Baxter solved the eight-vertex model in 1971, using a method based on
commuting transfer matrices, starting from a solution of what he then called
the generalized star-triangle equation, but what is now commonly called
the Yang--Baxter equation (YBE):
\begin{eqnarray}
&&\sum_{\alpha''}\sum_{\beta''}\sum_{\gamma''}
\omega{}^{\alpha''\beta''}_{\beta\,\,\,\alpha}(p,q)
\,\omega{}^{\gamma'\,\,\alpha'}_{\alpha''\gamma''}(q,r)
\,\omega{}^{\gamma''\beta'}_{\beta''\gamma}(p,r)\nonumber\\
&&\quad=\sum_{\alpha''}\sum_{\beta''}\sum_{\gamma''}
\omega{}^{\alpha'\,\,\beta'}_{\beta''\alpha''}(p,q)
\,\omega{}^{\gamma''\alpha''}_{\alpha\,\,\,\gamma}(q,r)
\,\omega{}^{\gamma'\beta''}_{\beta\,\gamma''}(p,r)
\label{eq8}
\end{eqnarray}
This equation is represented graphically in figure \ref{fig5}. From it one
can also derive a sufficient condition for the commutation of transfer
matrices and spin-chain Hamiltonians, generalizing work of McCoy and Wu who
had earlier initiated the search by showing that the general six-vertex model
transfer matrix commutes with a Heisenberg spin-chain Hamiltonian. To be more
precise, Baxter found that if $\omega{}^{\lambda\beta}_{\alpha\mu}=
\delta{}^{\lambda}_{\alpha}\delta{}^{\beta}_{\mu}$ for some choice of $p$
and $q$, some spin-chain Hamiltonians could be derived as logarithmic
derivatives of the transfer matrix.

\goodbreak\bigskip
\noindent{\mysfi Interaction-Round-a-Face model}
\par\noindent
Baxter introduced another language, namely that of the ``IRF-model" or
``interaction-round-a-face" model, which he introduced in connection
with his solution of the hard-hexagon model. This formulation is convenient
when studying one-point functions using the corner-transfer-matrix method.
\begin{figure}[tbh]
\begin{center}
\epsfig{file=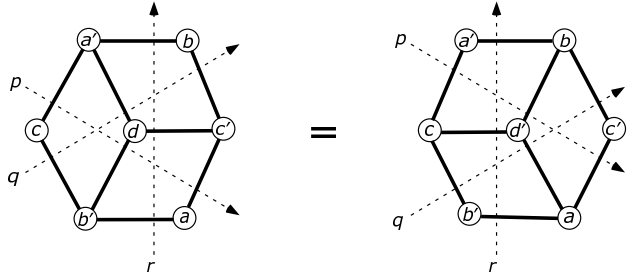,width=0.6\hsize}
\end{center}
\caption{IRF model Yang--Baxter equation.}
\label{fig7}
\end{figure}
Now the integrability condition can be represented graphically as in figure
\ref{fig7} or algebraically as
\begin{eqnarray}
&&\sum_{d}
w{}^{a'd}_{c\,\,b'}(p,q)\,w{}^{a'b}_{d\,\,c'}(q,r)\,w{}^{d\,\,c'}_{b'a}(p,r)
\nonumber\\
&&\qquad=\sum_{d'}
w{}^{b\,\,c'}_{d'a}(p,q)\,w{}^{c\,\,d'}_{b'a}(q,r)\,w{}^{a'b}_{c\,\,d'}(p,r)
\label{eq9}
\end{eqnarray}
The spins live on faces enclosed by rapidity lines and the weights
$w{}^{dc}_{ab}(p,q)$ are assigned as in figure \ref{fig6}.

Baxter discovered a new principle based on eqs.\ \ref{eq8} and \ref{eq9},
which he called $Z$-invariance as it expresses an invariance of the partition
function $Z$ under moves of rapidity lines. This also implies that typical
one-point functions are independent of the values of the rapidities, while
two-point functions can only depend on the values of the rapidities of
rapidity lines crossing between the two spins considered. Many recent results
on correlation functions in integrable models depend on this observation of
Baxter.

\goodbreak\bigskip
\noindent{\mysfi IRF-vertex model}
\par\noindent
In figure \ref{fig6}, we have also defined mixed IRF-vertex model weights
$W{}^{\lambda\beta}_{\alpha\mu}|{}^{dc}_{ab}(p,q)$. (We could
put further state variables on the vertices, but then the natural thing to do
is to introduce new effective weights summing over the states at each vertex.)
With the choice made
\begin{figure}[tbh]
\begin{center}
\epsfig{file=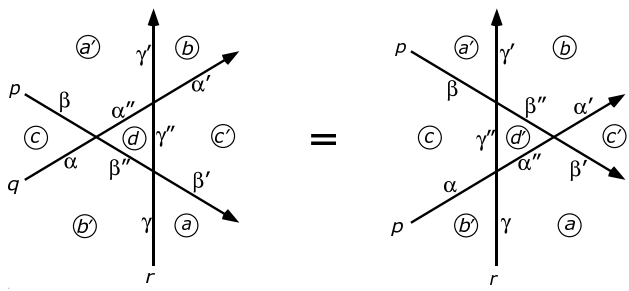,width=0.6\hsize}
\end{center}
\caption{General Yang--Baxter equation.}
\label{fig8}
\end{figure}
a more general Yang--Baxter equation can be represented as in figure
\ref{fig8}, or by
\begin{eqnarray}
&&\sum_{\alpha''}\sum_{\beta''}\sum_{\gamma''}\sum_{d}
W{}^{\alpha''\!\beta''}_{\beta\,\,\alpha}\!|{}^{a'd}_{c\,b'}(p,q)\cr
&&\hspace{6em}\times
\,W{}^{\gamma'\,\alpha'}_{\alpha''\gamma''}\!|{}^{a'b}_{d\,c'}(q,r)
\,W{}^{\gamma''\beta'}_{\beta''\gamma}\!|{}^{d\,c'}_{b'a}(p,r)\nonumber\\
&&=\quad\sum_{\alpha''}\sum_{\beta''}\sum_{\gamma''}\sum_{d'}
W{}^{\alpha'\,\beta'}_{\beta''\alpha''}\!|{}^{b\,c'}_{d'a}(p,q)\cr
&&\hspace{6em}\times
\,W{}^{\gamma''\!\alpha''}_{\alpha\,\,\gamma}\!|{}^{c\,d'}_{b'a}(q,r)
\,W{}^{\gamma'\beta''}_{\beta\,\gamma''}\!|{}^{a'b}_{c\,d'}(p,r)
\label{eq10}
\end{eqnarray}

\goodbreak\bigskip
\noindent{\mysfi Quantum Inverse Scattering Method}
\par\noindent
The Leningrad school of Faddeev incorporated the methods of Baxter and Yang
in their so-called {\em Quantum Inverse Scattering Method} (QISM), coining
the term {\em Quantum Yang--Baxter Equations} (QYBE) for the equations
[\ref{eq8}]. If special limiting values of $p$ and $q$ can be found, say as
$\hbar\to0$, such that $\omega{}^{\lambda\beta}_{\alpha\mu}=
\delta{}^{\lambda}_{\mu}\delta{}^{\beta}_{\alpha}+{\rm O}(\hbar)$, one can
reduce [\ref{eq8}] to the {\em Classical Yang--Baxter Equations} (CYBE) by
expanding up to the first nontrivial order in expansion variable $\hbar$.
These determine the integrability of certain models of classical mechanics by
the inverse scattering method and the existence of Lax pairs.

\goodbreak\bigskip
\noindent{\mysfb Checkerboard generalizations}
\par\noindent
Star-triangle equations [\ref{eq3}] and [\ref{eq4}] imply that there are
further generalizations of the Yang--Baxter equations, namely those for which
the faces enclosed by the rapidity lines are alternatingly colored black and
\begin{figure}[tbh]
\begin{center}
\epsfig{file=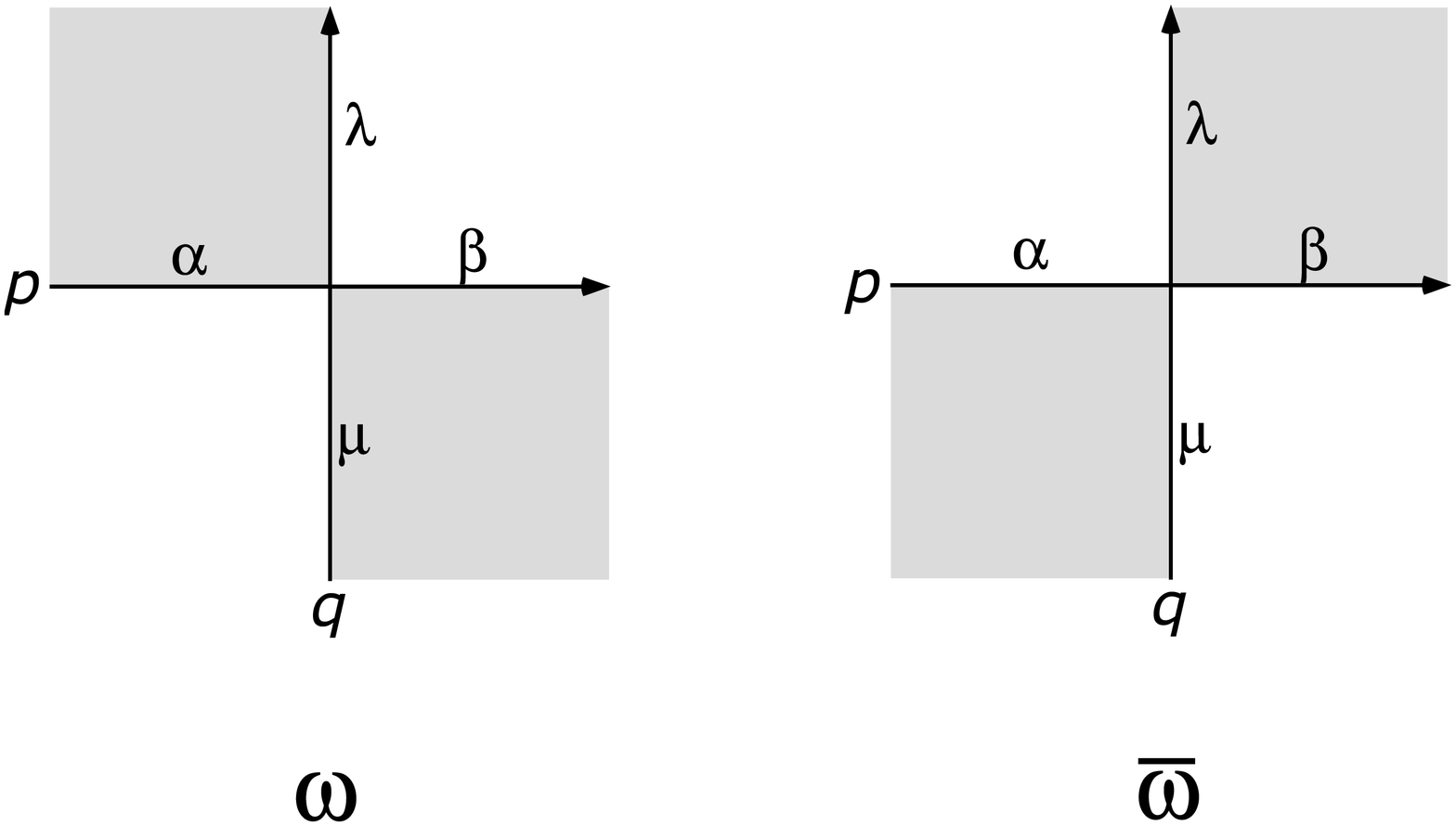,width=0.5\hsize}
\end{center}
\begin{center}
\epsfig{file=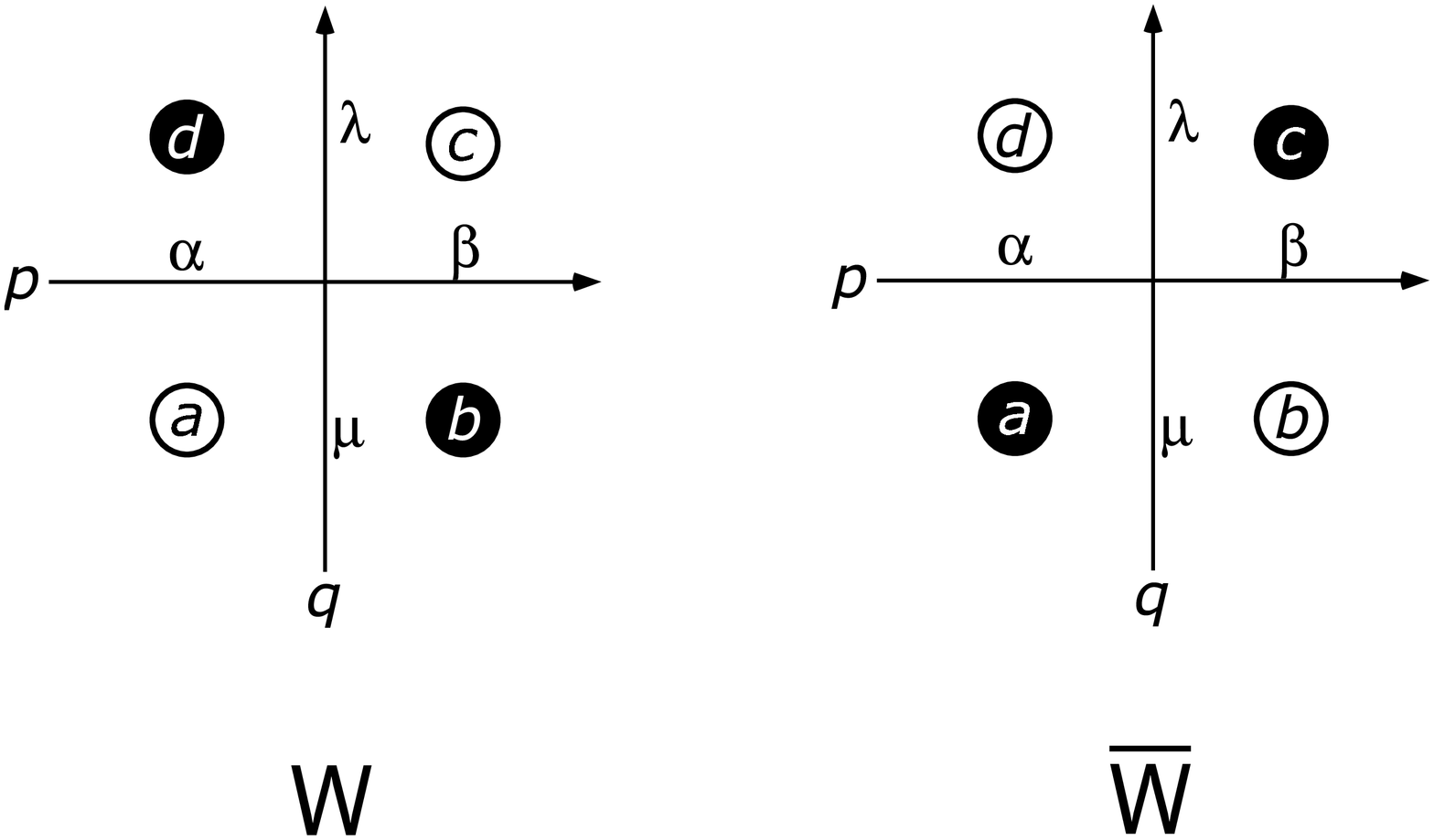,width=0.4\hsize}\hspace{0.1\hsize}
\epsfig{file=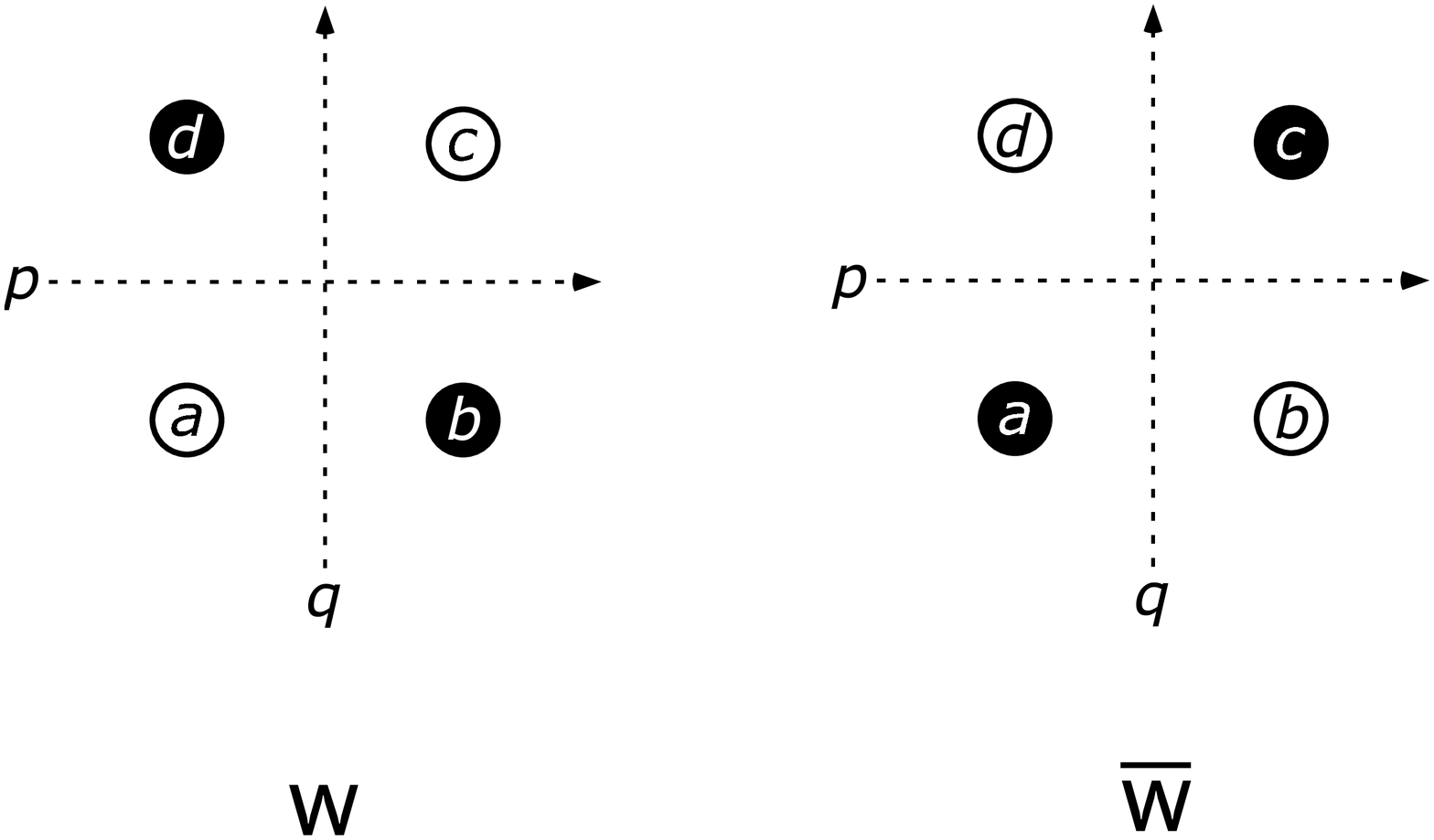,width=0.4\hsize}
\end{center}
\caption{Checkerboard versions of the weights.}
\label{fig9}
\end{figure}
white in a checkerboard pattern. We can then introduce either
vertex model weights $\omega{}^{\lambda\beta}_{\alpha\mu}(p,q)$
and $\overline\omega{}^{\lambda\beta}_{\alpha\mu}(p,q)$, or IRF-vertex
model weights $W{}^{\lambda\beta}_{\alpha\mu}|{}^{dc}_{ab}(p,q)$
and $\overline W{}^{\lambda\beta}_{\alpha\mu}|{}^{dc}_{ab}(p,q)$, or
IRF model weights $w{}^{dc}_{ab}(p,q)$ and $\overline w{}^{dc}_{ab}(p,q)$,
see figure \ref{fig9}.

The black faces are those where the spins of the spin model with weights
defined in figure \ref{fig2} live; the white faces are to be considered empty
in figures \ref{fig2} and \ref{fig3} (or, equivalently, they can be assumed
to host trivial spins that take on only a single value). Clearly, the
IRF-vertex model description contains all the other versions.

\goodbreak\bigskip
\noindent{\mysfi Checkerboard vertex model}
\par
\begin{figure}[tbh]
\begin{center}
\epsfig{file=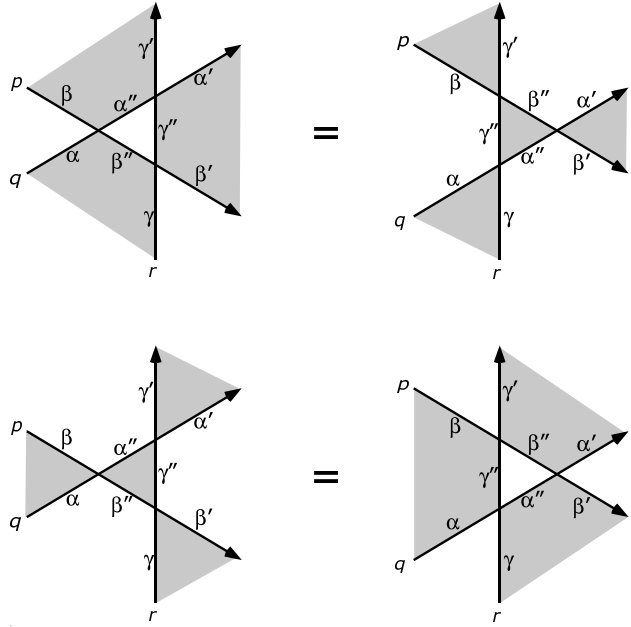,width=0.6\hsize}
\end{center}
\caption{Checkerboard vertex model Yang--Baxter equation.}
\label{fig10}
\end{figure}

\noindent
First we consider the checkerboard vertex model with weights
$\omega{}^{\lambda\beta}_{\alpha\mu}(p,q)$ and
$\overline\omega{}^{\lambda\beta}_{\alpha\mu}(p,q)$ as assigned in figure
\ref{fig9}. The YBE [\ref{eq8}] then generalizes to two sets of equations
\begin{eqnarray}
&&\sum_{\alpha''}\sum_{\beta''}\sum_{\gamma''}
\omega{}^{\alpha''\beta''}_{\beta\,\,\,\alpha}(p,q)
\,\omega{}^{\gamma'\,\,\alpha'}_{\alpha''\gamma''}(q,r)
\,\overline \omega{}^{\gamma''\beta'}_{\beta''\gamma}(p,r)\nonumber\\
&&\quad=R(p,q,r)\,
\sum_{\alpha''}\sum_{\beta''}\sum_{\gamma''}
\overline \omega{}^{\alpha'\,\,\beta'}_{\beta''\alpha''}(p,q)\cr
&&\hspace{8em}\times
\,\overline \omega{}^{\gamma''\alpha''}_{\alpha\,\,\,\gamma}(q,r)
\,\omega{}^{\gamma'\beta''}_{\beta\,\gamma''}(p,r)
\label{eq11}\\ \cr
&&\overline R(p,q,r)\,
\sum_{\alpha''}\sum_{\beta''}\sum_{\gamma''}
\overline \omega{}^{\alpha''\beta''}_{\beta\,\,\,\alpha}(p,q)\cr
&&\hspace{8em}\times
\,\overline \omega{}^{\gamma'\,\,\alpha'}_{\alpha''\gamma''}(q,r)
\,\omega{}^{\gamma''\beta'}_{\beta''\gamma}(p,r)\nonumber\\
&&\quad=\sum_{\alpha''}\sum_{\beta''}\sum_{\gamma''}
\omega{}^{\alpha'\,\,\beta'}_{\beta''\alpha''}(p,q)\cr
&&\hspace{8em}\times
\,\omega{}^{\gamma''\alpha''}_{\alpha\,\,\,\gamma}(q,r)
\,\overline \omega{}^{\gamma'\beta''}_{\beta\,\gamma''}(p,r)
\label{eq12}
\end{eqnarray}
where scalar factors $R$ and $\overline R$ have been added as in [\ref{eq3}]
and [\ref{eq4}]. These equations are represented graphically by figure
\ref{fig10}.

\goodbreak\bigskip
\noindent{\mysfi Checkerboard IRF model}
\par\noindent
The checkerboard IRF version of the YBE [\ref{eq8}] becomes
\begin{eqnarray}
&&\sum_{d}
w{}^{a'd}_{c\,\,\,b'}(p,q)
\,w{}^{a'b}_{d\,\,c'}(q,r)
\,\overline w{}^{d\,\,c'}_{b'a}(p,r)\nonumber\\
&&\quad=R(p,q,r)\,\sum_{d'}
\overline w{}^{b\,\,c'}_{d'a}(p,q)
\,\overline w{}^{c\,\,d'}_{b'a}(q,r)
\,w{}^{a'b}_{c\,\,d'}(p,r)
\label{eq13}\\ \cr
&&\overline R(p,q,r)\,\sum_{d}
\overline w{}^{a'd}_{c\,\,b'}(p,q)
\,\overline w{}^{a'b}_{d\,\,c'}(q,r)
\,w{}^{d\,\,c'}_{b'a}(p,r)\nonumber\\
&&\quad=\sum_{d'}
w{}^{b\,\,c'}_{d'a}(p,q)
\,w{}^{c\,\,d'}_{b'a}(q,r)
\,\overline w{}^{a'b}_{c\,\,d'}(p,r)
\label{eq14}
\end{eqnarray}
again with scalar factors $R$ and $\overline R$ added as in [\ref{eq3}] and
[\ref{eq4}]. These equations can
\begin{figure}[tbh]
\begin{center}
\epsfig{file=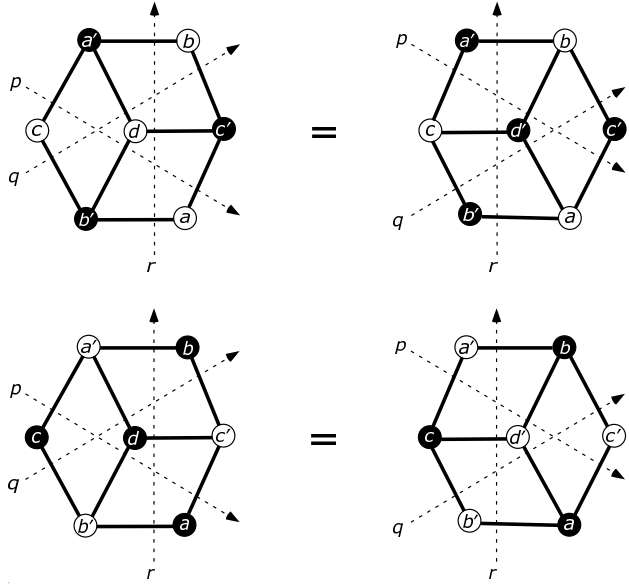,width=0.6\hsize}
\end{center}
\caption{Checkerboard IRF model Yang--Baxter equation.}
\label{fig11}
\end{figure}
now be represented graphically as in figure \ref{fig11}. Note that these
equations reduce to eqs.\ [\ref{eq3}] and [\ref{eq4}] if the spins on the
white faces are allowed to take only one value, which means that they can be
ignored.

\goodbreak\bigskip
\noindent{\mysfi Checkerboard IRF-vertex model}
\par
\begin{figure}[tbh]
\begin{center}
\epsfig{file=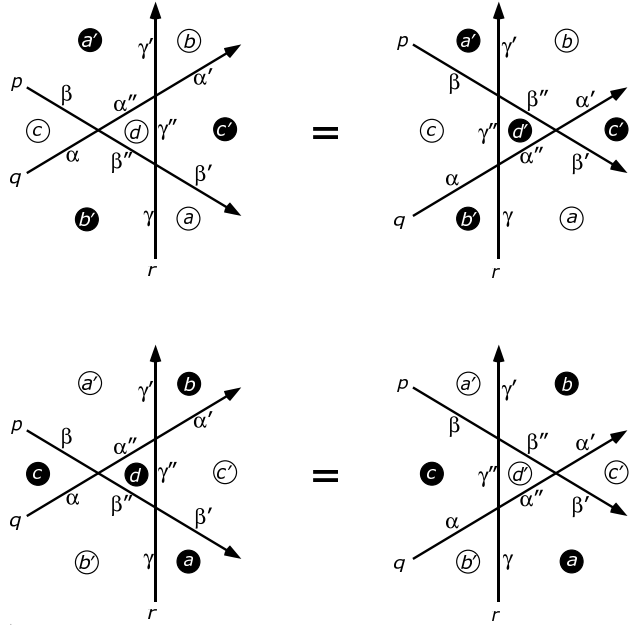,width=0.6\hsize}
\end{center}
\caption{Checkerboard Yang--Baxter equation.}
\label{fig12}
\end{figure}

\noindent
Finally, the most general case is represented by the checkerboard IRF-vertex
model, with weights defined in figure \ref{fig9}. For this case the YBE are
given by
\begin{eqnarray}
&&\sum_{\alpha''}\sum_{\beta''}\sum_{\gamma''}\sum_{d}
W{}^{\alpha''\!\beta''}_{\beta\,\,\alpha}\!|{}^{a'd}_{c\,b'}(p,q)
\cr&&\hspace{6em}\times\,
W{}^{\gamma'\,\alpha'}_{\alpha''\gamma''}\!|{}^{a'b}_{d\,c'}(q,r)
\,\overline W{}^{\gamma''\beta'}_{\beta''\gamma}\!|{}^{d\,c'}_{b'a}(p,r)
\nonumber\\
&&\quad=R(p,q,r)\,
\sum_{\alpha''}\sum_{\beta''}\sum_{\gamma''}\sum_{d'}
\overline W{}^{\alpha'\,\beta'}_{\beta''\alpha''}\!|{}^{b\,c'}_{d'a}(p,q)
\cr&&\hspace{6em}\times\,
\overline W{}^{\gamma''\!\alpha''}_{\alpha\,\,\gamma}\!|{}^{c\,d'}_{b'a}(q,r)
\,W{}^{\gamma'\beta''}_{\beta\,\gamma''}\!|{}^{a'b}_{c\,d'}(p,r)
\label{eq15}\\
&&\overline R(p,q,r)\,
\sum_{\alpha''}\sum_{\beta''}\sum_{\gamma''}\sum_{d}
\overline W{}^{\alpha''\!\beta''}_{\beta\,\,\alpha}\!|{}^{a'd}_{c\,b'}(p,q)
\cr&&\hspace{6em}\times\,
\overline W{}^{\gamma'\,\alpha'}_{\alpha''\gamma''}\!|{}^{a'b}_{d\,c'}(q,r)
\,W{}^{\gamma''\beta'}_{\beta''\gamma}\!|^{d\,c'}_{b'a}(p,r)\nonumber\\
&&\quad=\sum_{\alpha''}\sum_{\beta''}\sum_{\gamma''}\sum_{d'}
W{}^{\alpha'\,\beta'}_{\beta''\alpha''}\!|{}^{b\,c'}_{d'a}(p,q)
\cr&&\hspace{6em}\times\,
W{}^{\gamma''\!\alpha''}_{\alpha\,\,\gamma}\!|{}^{c\,d'}_{b'a}(q,r)
\,\overline W{}^{\gamma'\beta''}_{\beta\,\gamma''}\!|{}^{a'b}_{c\,d'}(p,r)
\label{eq16}
\end{eqnarray}
with its graphical representation in figure \ref{fig12}.

\goodbreak\bigskip
\noindent{\mysfb Formal equivalence of languages}\par\nobreak
\bigskip\nobreak
\noindent{\mysfi The square weight}\nobreak
\par\nobreak\noindent
Combining four weights of a checkerboard model in a square, as is done with
\begin{figure}[tbh]
\begin{center}
\epsfig{file=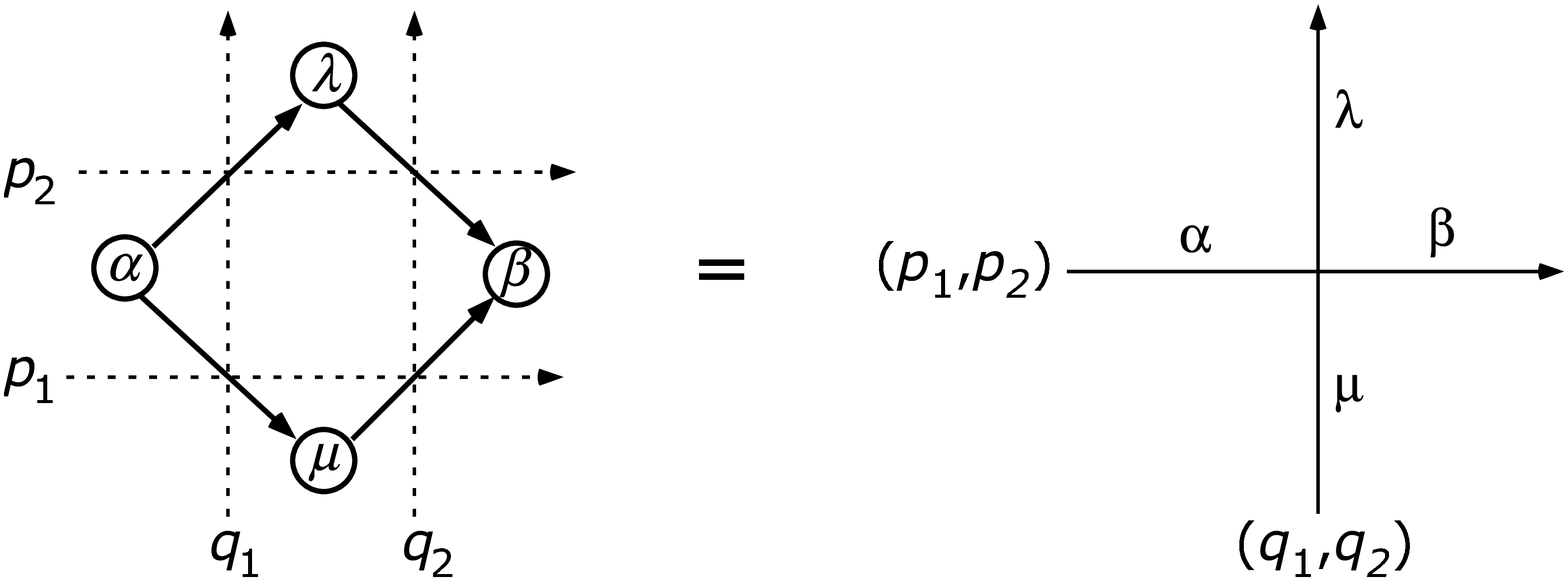,width=0.8\hsize}
\end{center}
\caption{Square weight as vertex weight.}
\label{fig13}
\end{figure}
four spin model weights in figure \ref{fig13}, we find a regular vertex model
weight with rapidities that are now pairs of the original ones. In formula,
this gives
\begin{eqnarray}
&W_{\alpha\mu}(p_1,q_1)\overline W_{\mu\beta}(p_1,q_2)
\overline W_{\alpha\lambda}(p_2,q_1)W_{\lambda\beta}(p_2,q_2)\cr
&=\,\omega{}^{\lambda\beta}_{\alpha\mu}(p_1,p_2;q_1,q_2)
\label{eq17}\end{eqnarray}

From any solution of [\ref{eq3}] and [\ref{eq4}] we can construct a solution
of YBE [\ref{eq8}] this way. This has been used by Bazhanov and Stroganov to
relate the integrable chiral Potts model with a cyclic representation of the
six vertex model.

\goodbreak\bigskip
\noindent{\mysfi Map to checkerboard vertex model}
\par\noindent
The checkerboard IRF-vertex model formulation contains all other versions
mentioned above as special cases. However, collecting the state variables in
triples, we can immediately translate it to a vertex model version, writing
\begin{eqnarray}
&&\omega{}^{\hat\lambda\hat\beta}_{\hat\alpha\hat\mu}(p,q)=
W{}^{\lambda\beta}_{\alpha\mu}|{}^{dc}_{ab}(p,q),\quad
\overline\omega{}^{\hat\lambda\hat\beta}_{\hat\alpha\hat\mu}(p,q)=
\overline W{}^{\lambda\beta}_{\alpha\mu}|{}^{dc}_{ab}(p,q)\nonumber\\
&&\qquad\qquad\mbox{if }\left\{\begin{array}{ll}
{\hat\lambda}=(d,\lambda,c),&{\hat\beta}=(b,\beta,c)\\
{\hat\alpha}=(a,\alpha,d),&{\hat\mu}=(a,\mu,b)
\end{array}\right.
\label{eq18}\\
&&\omega{}^{\hat\lambda\hat\beta}_{\hat\alpha\hat\mu}(p,q)=
\overline\omega{}^{\hat\lambda\hat\beta}_{\hat\alpha\hat\mu}(p,q)=
0\quad\mbox{otherwise}
\label{eq19}
\end{eqnarray}
In eq.\ [\ref{eq19}] we have set all vertex model weights zero that are
inconsistent with IRF-vertex configurations. Clearly, the translation of IRF
models and spin models to vertex models can be done similarly.

\goodbreak\bigskip
\noindent{\mysfi Map to spin model}
\par\noindent
We can furthermore translate each vertex model with weights assigned as in
figures \ref{fig6} or \ref{fig9} into a spin model with weights as in figure
\ref{fig2} by defining suitable spins in the black faces, after checkerboard
coloring. Each spin is then defined to be the ordered set of states on the
line segments of the vertex model, $\underline a=(\alpha_1,\alpha_2,\ldots)$,
ordering the line segments counter\-clockwise starting at, say, 12 o'clock.
We can then identify $\omega{}^{\lambda\beta}_{\alpha\mu}(p,q)=
W_{\underline a,\underline b}(p,q)$,
$\overline\omega{}^{\lambda\beta}_{\alpha\mu}(p,q)=
\overline W_{\underline a,\underline b}(p,q)$.
This is surely not very economical, as many of the weights will be equal, but
it helps show that all different versions of the checkerboard YBE are formally
equivalent.

Hence, we shall only use the vertex-model language in the following. It is
fairly straightforward to convert to the other formulations.

\goodbreak\bigskip
\noindent{\mysfb An sl(m$\boldsymbol{\vert}$n) example}
\par\noindent
One fundamental example is a $Q$-state model for which the rapidities have
$2Q+1$ components, $\vec p=(p_{-Q},\ldots,p_Q)$, $\vec q=(q_{-Q},\ldots,q_Q)$,
etc., and the states on the line segments are arranged in strings of
continuing conserved color. The vertex weights, for
$\alpha,\beta,\lambda,\mu=1,\ldots,Q$, are given by
\begin{equation}
\omega{}^{\lambda\beta}_{\alpha\mu}(\vec p,\vec q)=
\omega_0{}^{\lambda\beta}_{\alpha\mu}(p_0,q_0)\,
\frac{p_{+\lambda}\,q_{-\beta}}{q_{+\alpha}\,p_{-\mu}}
\label{eq20}\end{equation}
with
\begin{eqnarray}
\left\{
\begin{array}{l}
\omega_0{}^{\rho\rho}_{\rho\rho}(p_0,q_0)=\mathcal{N}\,
\sinh[\eta+\varepsilon_{\rho}(p_0-q_0)]\strut\\
\omega_0{}^{\rho\sigma}_{\sigma\rho}(p_0,q_0)=\mathcal{N}\,
G_{\rho\sigma}\sinh(p_0-q_0),\quad\rho\ne\sigma\strut^{\strut}\\
\omega_0{}^{\sigma\rho}_{\sigma\rho}(p_0,q_0)=\mathcal{N}\,
\mathrm{e}^{(p_0-q_0)\mathrm{sign}(\rho-\sigma)}\sinh\eta,
\quad\rho\ne\sigma\strut^{\strut}\\
\omega_0{}^{\lambda\beta}_{\alpha\mu}(p_0,q_0)=0,
\quad\mbox{otherwise}\strut^{\strut}
\label{eq21}\end{array}
\right.
\end{eqnarray}
where $\mathcal{N}$ is an arbitrary overall normalization factor and $\eta$ is
a constant. Furthermore, $\varepsilon_{\rho}=\pm1$ for $\rho=1,\ldots,Q$,
with
$m$ of them equal $+1$ and $n$ of them equal $-1$. The $G_{\rho\sigma}$'s are
constants satisfying $G_{\rho\sigma}=1/G_{\sigma\rho}$, which freedom is
allowed because the number of $\rho$-$\sigma$ crossings minus the number of
$\sigma$-$\rho$ crossings is fixed by the states on the boundary only, i.e.\
the choice of
$\alpha,\alpha',\beta,\beta',\gamma,\gamma'$ in YBE [\ref{eq8}] and figure
\ref{fig5}.

The solution [\ref{eq20}], [\ref{eq21}] has many applications. The case
$m\!=\!0$,
$n\!=\!2$ leads to the general six-vertex model; the $m\!=\!0$, $n\!=\!n$
case produces the fundamental intertwiner of affine quantum group
$\mathrm{U}_q\widehat{\mathfrak{sl}}(n)$, whereas the case $m\!=\!2$, $n\!=\!1$
corresponds to the supersymmetric one-dimensional $t$--$J$ model.

\goodbreak\bigskip
\noindent{\mysfb Operator formulations}
\par\noindent

\bigskip
\noindent{\mysfi The R-matrix}
\par\noindent
For a problem with $N$ rapidity lines, carrying rapidities $p_1,\ldots,p_N$,
we can introduce a set of matrices $\mathsf{R}_{ij}(p_i,p_j)$, for
$1\leqslant i<j\leqslant N$, with elements
\begin{equation}
R_{ij}(p_i,p_j){}^{\beta_1\ldots\beta_N}_{\alpha_1\ldots\alpha_N}=
\omega{}^{\beta_j\beta_i}_{\alpha_i\alpha_j}(p_i,p_j)
\prod_{k\ne i,j}\delta{}^{\beta_k}_{\alpha_k}
\label{eq22}
\end{equation}
In terms of these, the YBE [\ref{eq8}] can be rewritten in matrix form as
\begin{eqnarray}
&&\mathsf{R}_{jk}(p_j,p_k)
\mathsf{R}_{ik}(p_i,p_k)\mathsf{R}_{ij}(p_i,p_j)\cr
&&\qquad=\mathsf{R}_{ij}(p_i,p_j)\mathsf{R}_{ik}(p_i,p_k)
\mathsf{R}_{jk}(p_j,p_k)
\label{eq23}
\end{eqnarray}
where $1\leqslant i<j<k\leqslant N$.

\goodbreak\bigskip
\noindent{\mysfi The \v{R}-matrix}
\par\noindent
If we transpose the $\beta$ indices $\beta_i$ and $\beta_j$ in eq.\
[\ref{eq22}], we can define a set of matrices
${\check\mathsf{R}}_{i,i+1}(p,q)$ with elements
\begin{equation}
{\check R}_{i,i+1}(p,q){}^{\beta_1\ldots\beta_N}_{\alpha_1\ldots\alpha_N}=
\omega{}^{\beta_i,\beta_{i+1}}_{\alpha_i,\alpha_{i+1}}(p,q)
\prod_{k\ne i,i+1}\delta{}^{\beta_k}_{\alpha_k}
\label{eq24}
\end{equation}
Using these, the YBE [\ref{eq8}] can be rewritten in matrix form as
\begin{eqnarray}
&&{\check\mathsf{R}}_{i,i+1}(q,r){\check\mathsf{R}}_{i+1,i+2}(p,r)
{\check\mathsf{R}}_{i,i+1}(p,q)\cr
&&\qquad={\check\mathsf{R}}_{i+1,i+2}(p,q)
{\check\mathsf{R}}_{i,i+1}(p,r){\check\mathsf{R}}_{i+1,i+2}(q,r)
\label{eq25}
\end{eqnarray}
and
\begin{equation}
[{\check\mathsf{R}}_{i,i+1}(p,q),{\check\mathsf{R}}_{j,j+1}(r,s)]=0,
\quad\hbox{if}\quad \vert i-j\vert\ge 2
\label{eq26}
\end{equation}
In this formulation it is clear that many solutions can be found
``Baxterizing" Temperley--Lieb and Iwahori--Hecke algebras.

\goodbreak\bigskip
\noindent{\mysfb Classical Yang--Baxter Equations}
\par\noindent
If we expand
\begin{equation}
\mathsf{R}_{ij}(p_i,p_j)=\mathsf{1}+
\hbar\,\mathsf{X}_{ij}(p_i,p_j)+\mathrm{O}(\hbar^2)
\label{eq27}\end{equation}
in [\ref{eq23}] we get in second order in $\hbar$ the Classical Yang--Baxter
Equation (CYBE) as the vanishing of a sum of three commutators, i.e.
\begin{eqnarray}
&[\mathsf{X}_{ij}(p_i,p_j),\mathsf{X}_{ik}(p_i,p_k)]+
[\mathsf{X}_{ij}(p_i,p_j),\mathsf{X}_{jk}(p_j,p_k)]\cr
&\qquad+[\mathsf{X}_{ik}(p_i,p_k),\mathsf{X}_{jk}(p_j,p_k)]=0
\label{eq28}
\end{eqnarray}
introduced by Belavin and Drinfel'd, where $\mathsf{X}_{ij}$ is called
the classical $r$-matrix.

\goodbreak\bigskip
\noindent{\mysfb Reflection Yang--Baxter Equations}
\par\noindent
Cherednik and Sklyanin found a condition determining the solvability of
systems with boundaries, the Reflection Yang--Baxter Equations (RYBE), see
\begin{figure}[tbh]
\begin{center}
\epsfig{file=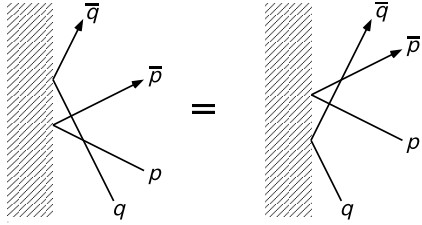,width=0.4\hsize}
\end{center}
\caption{Reflection Yang--Baxter Equation.}
\label{fig14}
\end{figure}
figure \ref{fig14}. Upon collisions with a left or right wall the
rapidity variable changes from $p$ to $\overline p$ and back. In most
examples, in which the rapidities are difference variables such that
$\mathsf{R}(p,q)=\mathsf{R}(p-q)$, one also has $\overline p=\mu-p$, with
$\mu$ some constant. The corresponding left boundary weights are
$K^{\beta}_{\alpha}(p,\overline p)$ satisfying
\begin{eqnarray}
&&{\check\mathsf{K}}_1(q,\overline q){\check\mathsf{R}}_{12}(\overline p,q)
{\check\mathsf{K}}_1(p,\overline p){\check\mathsf{R}}_{12}(q,p)\cr
&&\quad={\check\mathsf{R}}_{12}(\overline p,\overline q)
{\check\mathsf{K}}_1(p,\overline p)
{\check\mathsf{R}}_{12}(\overline q,p){\check\mathsf{K}}_1(q,\overline q)
\label{eq29}
\end{eqnarray}
with ${\check\mathsf{K}}_1(p,\overline p)$ defined by a direct product as in
[\ref{eq24}] appending unit matrices for positions $i\geqslant2$, and a
similar equation must hold for the right boundary. Most work has been done
for vertex models, while Pearce and coworkers wrote several papers on the
IRF-model version.

\goodbreak\bigskip
\noindent{\mysfb Higher dimensional generalizations}
\par\noindent
In 1980 Zamolodchikov introduced a three-dimensional generalization of
the YBE, the so-called Tetrahedron Equations (TE), and he found a
special solution. Baxter then succeeded in proving that this solution
satisfies all Tetrahedron Equations. Baxter and Bazhanov showed in 1992
that this solution can be seen as a special case of the sl($\infty$)
chiral Potts model. Several authors found further generalizations more
recently.

\goodbreak\bigskip
\noindent{\mysfb Inversion relations}
\par\noindent
When
$\omega{}^{\lambda\beta}_{\alpha\mu}(p,p)\propto
\delta{}^{\lambda}_{\alpha}\delta{}^{\beta}_{\mu}$, i.e.\ the weight
decouples when the two rapidities are equal, one can derive the local inverse
\begin{figure}[tbh]
\begin{center}
\epsfig{file=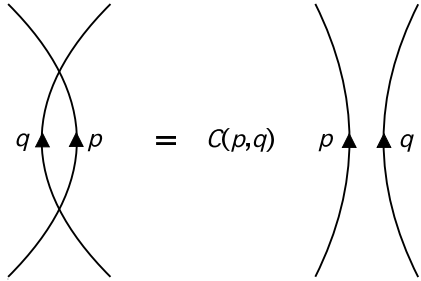,width=0.4\hsize}
\end{center}
\caption{Local inversion relation.}
\label{fig15}
\end{figure}
relation depicted in figure \ref{fig15}, which is a generalization of the
Reidemeister move of type II in figure \ref{fig4}. It is easily shown that
$C(q,p)=C(p,q)$.

\begin{figure}[tbh]
\begin{center}
\epsfig{file=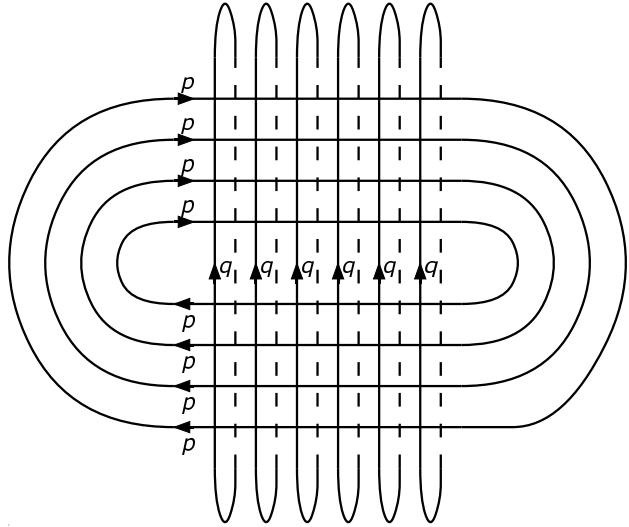,width=0.6\hsize}
\end{center}
\caption{Heuristic derivation of inversion relation.}
\label{fig16}
\end{figure}

This local relation implies also a global inversion relation which can be
found in many ways. The following heuristic way is the easiest: Consider the
situation in figure \ref{fig16}, with $N$ closed $p$-rapidity lines and $M$
closed $q$-rapidity lines. For $M$ and $N$ large, we may expect the partition
function of figure \ref{fig16} to factor asymptotically in top- and
bottom-half contributions. If each line segment carries a state variable that
can assume
$Q$ values, then the total partition function factors by repeated
application of the relation in figure \ref{fig15} into the contribution of
$M+N$ circles. Therefore,
\begin{equation}
Z=Q^{M+N}C(p,q)^{MN}\approx Z_{M,N}(p,q)Z_{N,M}(q,p)
\label{eq30}
\end{equation}
Taking the thermodynamic limit,
\begin{equation}
z(p,q)\equiv\lim_{M,N\to\infty}Z_{M,N}(p,q)^{1/MN}
\label{eq31}\end{equation}
one finds
\begin{equation}
z(p,q)z(q,p)=C(q,p)
\label{eq32}
\end{equation}
In many models eq.\ [\ref{eq32}], supplemented with some suitable symmetry and
analyticity conditions, can be used to calculate the free energy per site.

\goodbreak\bigskip
\noindent{\mysfb See also}
\par\noindent
\noindent{\mysfbs Bethe Ansatz, Quantum groups, Ising model, Potts model,
Vertex model, IRF model, Quantum inverse scattering method, Lax pair, Knot
theory, Temperley--Lieb algebra, Factorisable S-matrix}

\goodbreak\bigskip
\noindent{\mysfb Further Reading}
\begin{description}
\item{Au-Yang H and Perk JHH (1989) Onsager's star-triangle equation:
Master key to integrability.\ {\em Advanced Studies in Pure Mathematics}
19: 57--94.}

\item{Baxter RJ (1982) {\em Exactly Solved Models in Statistical Mechanics}.\
Academic Press, London.}

\item{Behrend RE, Pearce PA, O'Brien DL (1996) Interaction-Round-a-Face
Models with Fixed Boundary Conditions: The ABF Fusion Hierarchy. J.\
Stat.\ Phys.\ 84: 1--48.}

\item{Gaudin M (1983) {\em La Fonction d'Onde de Bethe}.\ Masson, Paris.}

\item{Jimbo M (ed) (1987) {\em Yang--Baxter Equation in Integrable Systems}.\
World Scientific, Singapore.}

\item{Kennelly AE (1899) The equivalence of triangles and three-pointed
stars in conducting networks.\ {\em Electrical World and Engineer}
34: 413--414.}

\item{Korepin VE, Bogoliubov NM, and Izergin AG (1993) {\em Quantum Inverse
Scattering Method and Correlation Functions}.\ Cambridge Univ. Press.}

\item{Kulish PP and Sklyanin EK (1981) Quantum spectral transform method.
Recent developments.\ In: Hietarinta J and Montonen C (eds)
{\em Integrable quantum field  theories}, Lecture Notes in Physics,
vol.\ 151, pp 61--119.\ Springer, Berlin.}

\item{Lieb EH and Wu FY (1972) Two-Dimensional Ferroelectric  Models.\ In:
Domb C and Green MS (eds) {\em Phase Transitions and Critical Phenomena},
vol.\ 1, pp 331--490.\ Academic Press, London.}

\item{Onsager L (1971) The Ising model in two dimensions.\ In: Mills RE,
Ascher E and Jaffee RI (eds) {\em Critical phenomena in alloys, magnets and
superconductors}, pp xix--xxiv, 3--12.\ McGraw-Hill, New York.}

\item{Perk JHH (1989) Star-triangle equations, quantum Lax pairs, and 
higher genus curves.\ {\em Proc. Symposia in Pure Math.} 49(1): 341--354.}

\item{Perk JHH and Schultz CL (1981) New families of commuting transfer
matrices in $q$-state vertex models. {\em Phys.\ Lett.} A84: 407--410.}

\item{Perk JHH and Wu FY (1986) Graphical approach to the nonintersecting
string model: star-triangle equation, inversion relation, and exact solution.\
{\em Physica} A138: 100--124.}

\item{Reidemeister K (1926) Knoten und Gruppen.\
{\em Abh.\ Math.\ Sem.\ Hamburg.\ Univ.} 5: 7--23.\
Elementare Begr\"undung der Knotentheorie.\
{\em ibid.} 24--32.}

\item{Yang CN (1967) Some exact results for the many-body problem
in one dimension with repulsive delta-function interaction.\
{\em Phys.\ Rev.\ Lett.} 19: 1312--1314.}

\item{Yang CN (1968) $S$ Matrix for the One-Dimensional $N$-Body Problem with
Repulsive or Attractive $\delta$-Function Interaction. {\em Phys.\ Rev.}
167: 1920--1923.}

\end{description}

Remark: most older work can be found from Jimbo's book above, while most
more recent work can be found at www.arXiv.org.

\goodbreak\bigskip\bigskip
\noindent{\mysfb Appendix: Relation of resistor networks with Gaussian model
and Potts model}
\par\noindent
We make next a few remarks on the relationship of resistor networks
with related Gaussian models and Potts models in certain limits.

We start with a graph $\rm G$ with vertices $j$ and edges
$\langle j,j'\rangle$. We associate with each vertex $j$ an electric
potential $\phi_j$ and with each edge $\langle j,j'\rangle$ a
resistance $R_{j,j'}$. According to Ohm's law the current from vertex
$j$ to a neighboring vertex $j'$ (i.e.\ $j'\in\,{\rm N\!b}_j$) is given by
\begin{equation}
I_{j,j'}=\frac{\phi_j-\phi_{j'}}{R_{j,j'}},
\label{eq33}
\end{equation}
whereas the power dissipated in the resistors is
\begin{equation}
P=\sum_{\langle j,j'\rangle}\frac{(\phi_j-\phi_{j'})^2}{R_{j,j'}},
\label{eq34}
\end{equation}
with a sum over all edges. Kirchhoff's second law is implicit in this
formulation. However, minimizing $P$ over the potential $\phi_j$ for all
internal vertices $j$ (i.e.\ $j\in\,{\rm Int_G}$), we find
\begin{equation}
\frac{\partial P}{\partial\phi_j}=0\quad\Longrightarrow\quad
\sum_{j'\in\,{\rm N\!b}_j}\frac{\phi_j-\phi_{j'}}{R_{j,j'}}=0
\quad\Longrightarrow\quad
\sum_{j'\in\,{\rm N\!b}_j}I_{j,j'}=0
\label{eq35}
\end{equation}
and Kirchhoff's first law emerges.

We can now define the associated Gaussian model on graph $\rm G$,
using $P$ for the interaction energy. The partition function is
\begin{equation}
Z=\bigg(\prod_{j\in\,{\rm Int_G}}\sqrt{\frac{\beta}{\pi}}
\int_{-\infty}^{\infty}{\rm d}\phi_j\bigg)\,
\exp\bigg(-\beta\sum_{\langle j,j'\rangle}
\frac{(\phi_j-\phi_{j'})^2}{R_{j,j'}}\bigg),
\label{eq36}
\end{equation}
which depends on the ``potentials" of the exterior vertices.
In the zero-temperature limit $\beta\to\infty$ the free energy
tends to $\min P$, defining the equilibrium state of the resistor network.

We can even go further and derive a star-triangle relation for the
Gaussian model. Doing a single Gaussian integration over $\phi_0$
at the internal point 0 of the star with vertices $j=0,1,2,3$, we find
\begin{eqnarray}
&&\sqrt{\frac{\beta}{\pi}}\int_{-\infty}^{\infty}{\rm d}\phi_0\,
\exp\bigg(-\beta\sum_{j=1}^3\frac{(\phi_j-\phi_0)^2}{R_{j,0}}\bigg)
\nonumber\\
&&=\Bigg(\prod_{j=1}^3\frac{R_{j,0}^{\;1/3}}{R_{j,j+1}^{\;1/6}}\Bigg)\,
\exp\bigg(-\beta\sum_{j=1}^3\,\frac{(\phi_j-\phi_{j+1})^2}
{R_{j,j+1}}\bigg)
\label{eq37}
\end{eqnarray}
with $j+1\equiv1$ for $j=3$ and $R_{j,j+1}$ defined by
\begin{equation}
\frac{R_{j,j+1}}{R_{j,0}R_{j+1,0}}=\sum_{k=1}^3\frac{1}{R_{k,0}},
\quad\hbox{for }j=1,2,3.
\label{eq38}
\end{equation}
The front factor in the RHS of [\ref{eq37}] has been factorized this way
using the product of [\ref{eq38}] over $j=1,2,3$. It is related to the
scalar factors $R(p,q,r)$ and ${\overline R}(p,q,r)$ in [\ref{eq3}] and
[\ref{eq4}]. The continuous variables $\phi_j$ correspond
to the discrete state variables $a$, $b$, $c$, $d$ and the integral
$\sqrt{\beta/\pi}\int{\rm d}\phi_0$ to the sum $\sum_d$ there.

In the zero-temperature limit $\beta\to\infty$ the integral in the LHS of
[\ref{eq37}] is dominated by the maximum of the integrand and the RHS
is also dominated by the exponential. The star-triangle relation
[\ref{eq1}], [\ref{eq2}] emerges, when identifying $R_{j,0}=Z_j$,
($j=1,2,3)$, and
$R_{1,2}={\overline Z}_3$, $R_{2,3}={\overline Z}_1$,
$R_{3,1}={\overline Z}_2$ in [\ref{eq38}].

Another mapping has been given by Fortuin and Kasteleyn, who have shown
that any planar resistor network is related to an $N\to0$ limit of a
corresponding $N$-state Potts model [See, e.g.,  Section IV$\,$C of F.Y.~Wu,
Rev.\ Mod.\ Phys.\ 54, 235--268 (1982)]. Given Potts interaction energy
\begin{equation}
\mathcal{E}=-\sum_{\langle j,j'\rangle} J_{j,j'}
\delta_{\sigma_{j\vphantom{j'}},\sigma_{j'\vphantom{j'}}},
\label{eq39}
\end{equation}
we make, for each edge $r=\langle j,j'\rangle$, the special combinations
\begin{equation}
x_r=\frac{{\rm e}^{\beta J_r}-1}{\sqrt{N}},\quad
\bar{x}_r=\frac{{\rm e}^{\beta\bar{J}_r}-1}{\sqrt{N}}
\equiv\frac{1}{x_r}.
\label{eq40}
\end{equation}
The last equation defines the corresponding coupling constant
$\bar{J}_r$ of the dual Potts model.
The limit $N\to0$ is defined via the six-vertex model and in this limit
we may choose $\beta=\sqrt{N}$ and find
\begin{equation}
x_r=J_r=c/R_r,\quad \bar{x}_r=\bar{J}_r=c/\overline{R}_r,
\label{eq41}
\end{equation}
with $c$ some positive constant.

Writing the star-triangle equation [\ref{eq3}] for the $N$-state Potts model
in terms of $x$ and $\bar{x}$ with
\begin{equation}
W_{ab}(p,q)\!=\!1+\sqrt{N}x(p-q)\delta_{a,b},\quad
\overline{W}_{ab}(p,q)\!=\!1+\sqrt{N}\bar{x}(p-q)\delta_{a,b},\quad
\label{eq42}
\end{equation}
we find five independent equations. Eliminating $R(p,q,r)$, we arrive at
\begin{eqnarray}
&\bar{x}(p\!-\!q)x(p\!-\!r)\bar{x}(q\!-\!r)
=\bar{x}(p\!-\!q)+x(p\!-\!r)+\bar{x}(q\!-\!r)+\sqrt{N},&\hspace{-1em}
\nonumber\\
&\hspace{-10em}\hbox{and}\quad x(u)\bar{x}(u)\equiv1,&\hspace{-10em}
\label{eq43}
\end{eqnarray}
with the well-known solution
$x(u)=\sin(u)/\sin(\theta\!-\!u)=1/\bar{x}(u)$, where
$\theta\equiv\arccos(\sqrt{N}/2).$
In the limit $N\to0$ this solution becomes $x(u)=\tan(u)$,
$\bar{x}(u)=\cot(u)$, while, in view of [\ref{eq41}], Eq.~[\ref{eq43}]
reduces to [\ref{eq1}] and [\ref{eq2}] with
\begin{eqnarray}
&&Z_1=c\tan(p\!-\!q),\quad Z_2=c\cot(p\!-\!r),\quad
Z_3=c\tan(q\!-\!r),\nonumber\\
&&\overline{Z}_1=c\cot(p\!-\!q),\quad \overline{Z}_2=c\tan(p\!-\!r),\quad
\overline{Z}_3=c\cot(q\!-\!r).
\label{eq44}
\end{eqnarray}
This is a rapidity parametrization of the solution of [\ref{eq1}]
and [\ref{eq2}].
\end{document}